\documentstyle[aps,twocolumn,epsf,psfig]{revtex}

\begin{document}
\draft

\title {The breakdown of adiabatic polaron theory in one, two, and three dimensions\\
and the reformation of the large polaron concept}

\author{Aldo~H.~Romero${^{1,3}}$, David W. Brown${^2}$ and Katja Lindenberg${^3}$}

\address
{${^1}$
Department of Physics,\\
University of California, San Diego, La Jolla, CA 92093-0354}

\address
{${^2}$
Institute for Nonlinear Science,\\
University of California, San Diego, La Jolla, CA 92093-0402}

\address
{${^3}$
Department of Chemistry and Biochemistry,\\
University of California, San Diego, La Jolla, CA 92093-0340} 

\date{\today} 

\maketitle

\begin{abstract}

We prove variationally that at weak coupling in one, two, and three dimensions there exist correlated electron-phonon states below the approximate ground states characteristically found by adiabatic polaron theory.
Besides differing non-trivially in quantitative aspects such as the value of the ground state energy, these improved ground states are found to differ significantly in qualitative aspects such as correlation structure and scaling behavior.
These differences are sufficiently severe as to require a reevalutation of the physical meaning attached to such widely used terms as "large polaron" and "self-trapping transition".

\end{abstract}
%\vspace{0.25in}

\pacs{PACS numbers: 71.38.+i, 71.15.-m, 71.35.Aa, 72.90.+y}

\narrowtext

The theory of polarons in condensed matter is a venerable subject that has drawn the attention of many creative minds applying an impressive array of theoretical techniques for more than half of this century.
In more recent days, there has been a surge of activity in this area driven in part by a revitalized interest in the potential role of polarons in superconductivity and by some pivotal advances in computational techniques.

Although there is much now that can be done to reveal the secrets polarons have held close for many years, many of the concepts shaping the perspective from which these are viewed are decades old.
It should not be surprising that some updating might be required.

One such concept is the dichotomy of "free" vs. "self-trapped" states and the intimately related issues of the dependence of polaron structure on lattice dimensionality.
This concept arises from adiabatic theory \cite{Toyozawa61,Toyozawa63,Emin73,Sumi73,Emin76,Ueta86,Kabanov93,Toyozawa80a,Silinsh94,Song96,Toyozawa97}, which we use rather broadly to include a number of semi-classical approaches and uncontrolled approximations that may not explicitly invoke the adiabatic approximation in an obvious way, but which are essentially indistinguishable in their result.
According to adiabatic theory, the minimum energy state of an electron-phonon system of any dimensionality at sufficiently strong coupling is a compact, spatially-localized state.
With decreasing electron-phonon coupling strength, however, the energy of this branch of localized solutions is found to rise until, for $D \ge 2$, it penetrates the free-electron energy band at some finite coupling strength characteristic of the particular lattice structure and type of electron-phonon interaction.
On the weak-coupling side of this crossover, the minimum energy states are found to be the free electron states; hence the notion of a self-trapping transition arises as a crossover in the global ground state from "free" to "self-trapped" character.
In 1-D, the minimum-energy solution at any value of electron-phonon coupling is found to be a localized state; hence, the notion that all 1-D electron-phonon states are "self-trapped".

Although appealing in many ways, very widely accepted, and correct in certain respects, almost every "interesting" element of this adiabatic characterization of polaron structure and self-trapping is here proven to be break down in the weak-coupling regime, {\it even in 1-D}.
This breakdown of adiabatic theory can be demonstrated with a simple and now decades old variational method attributable to Merrifield \cite{Merrifield64,Zhao97a}.
Although the Merrifield method is not the method of choice for wide-ranging, high precision calculation, its use for our present purpose lends some transparency to the demonstration since it is quite closely related to some very typical adiabatic approaches while simultaneously comparing favorably with more accurate methods \cite{Romero98a}.

We use the Holstein Hamiltonian \cite{Holstein59a,Holstein59b}
\begin{eqnarray}
\hat{H}&=& - J \!\!\! \sum_{ < \vec{m} , \vec{n} > } \! a_{\vec{m}}^{\dagger} a_{\vec{n}}
+ \hbar \omega \sum_{\vec{n}} b_{\vec{n}}^{\dagger} b_{\vec{n}} \nonumber \\
&& ~~~ - g \hbar \omega \sum_{\vec{n}} a_{\vec{n}}^{\dagger}
a_{\vec{n}} ( b_{\vec{n}}^{\dagger} + b_{\vec{n}} ) ~,
\label{eq:holstein}
\end{eqnarray}
in which $a_{\vec{n}}^\dagger$ creates a single electron in the rigid-lattice Wannier state at site ${\vec{n}}$, and $b_{\vec{n}}^\dagger$ creates a quantum of vibrational energy $\hbar \omega$ in the Einstein oscillator at site ${\vec{n}}$.
All sums run over the entire $D$-dimensional lattice of edge-length $N$; the restriction $< \vec{m} , \vec{n} >$ limits $\vec{m}$ and $\vec{n}$ to be nearest neighbors along the each crystal axis, and $J$ is the electron transfer integral between such neighboring sites.
We note that throughout this paper, "coupling strength" refers to the dimensionless local coupling parameter $g$, and not the relative measure $\lambda = g^2 \hbar\omega/2J$ common in adiabatic approaches.

In one very characteristic approach, the ground state of the electron-phonon system is approximated by a Pekar form
\begin{equation}
| \psi \rangle = | \alpha \rangle \otimes | \beta \rangle ~,
\label{eq:pekar}
\end{equation}
in which $| \alpha \rangle$ is a one-electron state and $| \beta \rangle$ represents the (multi-phonon) state of the lattice.
The lattice state $| \beta \rangle$ is sometimes subsumed into a classical treatment of the lattice; however, since the lowest-energy lattice states consistent with arbitrary classical configurations are coherent states, a very typical choice for the components of the Pekar form are
\begin{eqnarray}
| \alpha \rangle &=& \sum_{\vec{n}} \alpha_{\vec{n}} a_{\vec{n}}^{\dagger} |0 \rangle ~,
\label{eq:alpha} \\
| \beta \rangle &=& \exp \left[ \sum_{\vec{n}} ( \beta_{\vec{n}} b_{\vec{n}}^{\dagger} - \beta_{\vec{n}}^* b_{\vec{n}} ) \right] ~.
\label{eq:beta}
\end{eqnarray}
Applying such a state to the model Hamiltonian and then varying the parameters $\{ \alpha , \beta \}$ to minimize the energy yields the adiabatic character outlined above.
That is, in all of 1-D and at sufficiently strong coupling in 2-D and 3-D, a balance can be struck between the spread of the electronic component ($\alpha$) and the focussing tendencies of the lattice component ($\beta$) to yield non-trivially localized ground states, while at weak coupling in 2-D and 3-D, the free electron ground state is energetically favored.

Rather than account for electronic spreading within the Pekar form, the Merrifield method uses a delocalized state that strictly satisfies the Bloch condition regardless of coupling regime
\begin{eqnarray}
| \Psi ( \vec{\kappa} ) \rangle &=& \frac 1 {\sqrt{N^D}} \sum_{\vec{m}} \left\{
e^{i \vec{\kappa} \cdot \vec{m} } a_{\vec{m}}^{\dagger} \right. \nonumber \\
&& ~~~~~ ~~~~~ \left. \times \exp \left[ \sum_{\vec{n}}
( \beta^{\vec{\kappa} *}_{\vec{n} - \vec{m}} b_{\vec{n}}^{\dagger}
- \beta^{\vec{\kappa} *}_{\vec{n} - \vec{m}} b_{\vec{n}} ) \right] \right\} ~.
\end{eqnarray}
This state can be obtained from (\ref{eq:beta}) by first localizing the electronic component (setting $\alpha_n = \delta_{n0}$) and then forming the appropriate superposition of displaced replicas ($\vec{\kappa} = 0$ for the global ground state).

As with the Pekar form, the Merrifield state is applied to the system Hamiltonian and the parameters $\{ \beta \}$ are varied to minimize the energy.
The resulting self-consistency equations are then solved numerically.
Our calculations were performed in nearly-real time on a single-processor Sun Microsystems Sparc 10 desktop workstation.
Although for the parameter values considered here convergence to "bulk" values is accomplished on modest-sized lattices, the data presented in this paper were all obtained on isotropic cubic lattices of edge length $N=64$ in order to obviate any potential concerns relating to boundary effects; thus, in 1-D, 2-D, and 3-D, $N^D = 64$, $4096$, and $262,144$ respectively.

Although the Merrifield method is {\it not} accurate at intermediate and strong coupling (its distorts the self-trapping transition and converges toward the {\it first} order of strong-coupling perturbation theory, missing the important second-order correction \cite{Alexandrov95}), it has the virtue of providing a straightforward variational confirmation of weak-coupling perturbation theory.
This is significant because the nonlinear nature of the stability arguments supporting the adiabatic characterization of polaron structure naturally raise some concerns over the reliability of weak-coupling perturbation theory that are hereby put to rest.

\begin{figure*}[htb]
\begin{center}
\leavevmode
\epsfysize = 4.25in
\epsfxsize = 3.5in
\epsffile{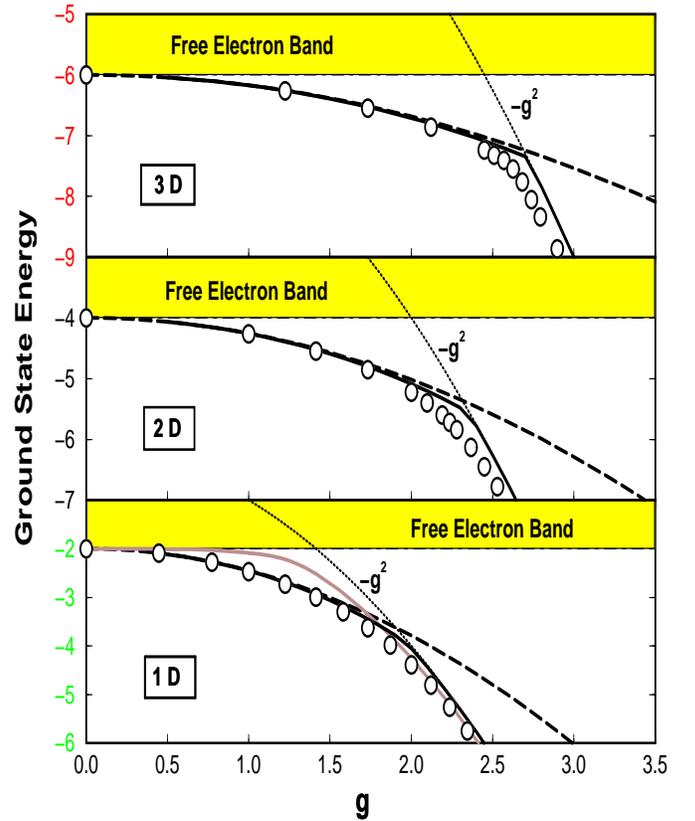}
\end{center}
\caption
{Ground state energy in units of $\hbar \omega$ in 1-D, 2-D, and 3-D for $J = \hbar \omega$.
Chain-dotted line:  Bottom of the free electron band at $-2DJ$.
Dotted line:  Strong-coupling asymptote, $E(0) \sim -g^2$.
Dashed line:  Weak-coupling perturbation theory.
Bold solid line:  Merrifield method.
Faint solid line:  Adiabatic theory on a discrete 1-D lattice; data kindly provided by G. Kalosakas \protect\cite{Kalosakas98,Kalosakas97}\protect.
Points:  Quantum Monte Carlo, data kindly provided by P. E. Kornilovitch \protect\cite{Kornilovitch98a,Kornilovitch98b}\protect.
}
\label{fig:ground}
\end{figure*}

\begin{figure*}[htb]
\begin{center}
\leavevmode
\epsfxsize = 3.5in
\epsffile{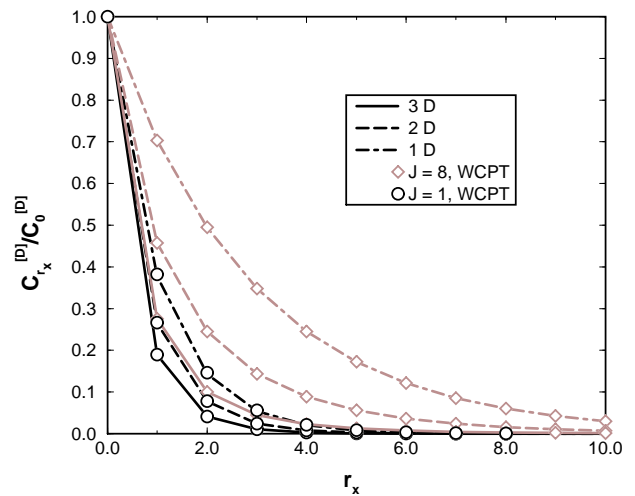}
\end{center}
\caption
{Correlation decay along the $x$ axis in 1-D, 2-D, and 3-D, as measured by
${C}_{\vec{r}}^{[D]} / {C}_{0}^{[D]} $.
Symbols indicate the results of weak-coupling perturbation theory.
Segmented curves indicate the results of the Merrifield method for $g = 1/2$.
Solid lines:  3-D.  Chain-dotted lines 2-D.  Dashed lines:  1-D.
Bold lines, circles:  $J=1$.
Faint lines, diamonds:  $J=8$.}
\label{fig:corr123d}
\end{figure*}

As the comparisons in Figure~\ref{fig:ground} show, the Merrifield method provides variational proof that in the weak-coupling regime:
i) the free electron states are {\it not} the lowest energy states {\it in any dimension},
2) the "large polaron" states of 1-D adiabatic theory are {\it not} the lowest energy states in 1-D, and
3) the correlated electron-phonon states that {\it do} minimize the energy are more consistent with weak-coupling perturbation theory than with adiabatic theory.
This consistency between weak-coupling perturbation theory and the Merrifield method is not limited to the ground state energy, but extends to the
polaron effective mass \cite{Romero98e},
kinetic energy \cite{Romero98c},
and the spatial structure of the electron-phonon correlations as well \cite{Romero98d};
examples of the latter are shown in Figure~\ref{fig:corr123d}.

We are thus forced to consider the crucial question of just how to characterize the correlated electron-phonon ground states of the weak-coupling regime.
For historical reasons at the heart of this paper, there has been considerable reluctance to characterize these states as "large polarons", though in every dimension these are the weak-coupling complement to the strong-coupling states that are unambiguously and universally characterized as "small polarons".
The principal reason for this reluctance appears to be that "large polaron" has come to be identified with the archetypical finding of adiabatic theory in 1-D that the minimum-energy states at weak-coupling are broad, pulse-shaped structures characterized by a locking relation between the electron and phonon amplitudes (e.g., $\beta_n = g | \alpha_n |^2$ when using (\ref{eq:holstein}) - (\ref{eq:beta})) and by certain scaling properties in the continuum limit.
The Pekar form (\ref{eq:pekar})-(\ref{eq:beta}) in particular applied to (\ref{eq:holstein}) at sufficiently weak coupling (permitting the continuum limit) yields
\begin{eqnarray}
\alpha(x) &=& \left( \frac{\lambda}{2 l} \right) ^{1/2} {\rm sech} ( \lambda x / l) ~, \\
\beta(x) &=& \left( \frac{\lambda g}{2} \right) {\rm sech}^2 (\lambda x /l) ~,
\end{eqnarray}
in which $\lambda = g^2 \hbar\omega/2J$ and $l$ is the lattice constant.

Yet, the results displayed in Figure~\ref{fig:ground} prove that "large polaron" ground states as here described do not exist in 1-D for the same reasons that "free electron" ground states do not exist in 2-D and 3-D; viz., the 1-D "large polarons" of adiabatic theory lie at higher energy than the correlated electron-phonon states indicated with great mutual consistency by weak-coupling perturbation theory, band-theoretic variational methods, quantum Monte Carlo, and others \cite{Romero98a}.

This disagreement is not a casual matter of complex correlations being described with fewer or greater parameters, or exponential vs. gaussian tails, or differing numbers of retained orders that ultimately are of little consequence when handled with appropriate perspective and care.
Rather, the adiabatic notion of the "large polaron" in 1-D differs from actual weak-coupling polaron structure in the most essential of polaron properties, the polaron size and in the scaling of this size with system parameters.
For example, consider the correlation function
\begin{equation}
C_{\vec{r}}^{[D]} = \langle \hat{C}_{\vec{r}}^{[D]} \rangle = \frac 1 {2g} \langle \sum_{\vec{n}} a_{\vec{n}}^{\dagger} a_{\vec{n}} (b_{\vec{n}+\vec{r}}^{\dagger} + b_{\vec{n}+\vec{r}} ) \rangle  ~.
\label{eq:corrfunc}
\end{equation}
that can be viewed as measuring the shape of the polaron lattice distortion around the instantaneous position of the electron, or, essentially equivalently in view of the strongly local character of the electron-phonon coupling, as an image of the electron density one associates with a localized polaron.

In 1-D, using the collected results following from the Pekar form, the spatial variance of this correlation function is straightforwardly found to be
\begin{equation}
\sigma^2 = \frac {\pi ^2} 3 \left( \frac {J} {g^2 \hbar \omega} \right) ^2 ~,
\label{eq:radius}
\end{equation}
which we note diverges strongly at weak coupling.
On the other hand, the result that follows analytically from weak coupling perturbation theory in any dimension \cite{Romero98d} is
\begin{equation}
\sigma^2_{ij} \equiv \sum_{\vec{r}} r_i r_j {C}_{\vec{r}}^{[D]} = \delta_{ij} \frac {2J} {\hbar \omega} ~.
\end{equation}
which is independent of $g$ and scales differently with $J/\hbar\omega$ than does the adiabatic result.
This result is confirmed at weak-coupling by the Merrifield data in 1-D, 2-D, and 3-D (see Figure~\ref{fig:corr123d}), and by data from the more accurate Global-Local method in 1-D (higher D pending).

For the cases illustrated in Figure~\ref{fig:corr123d}, the "large polarons" of adiabatic theory would be considerably wider; based on the square root of the ratio of the variances, about 5 times wider in the $J=1$ case and 14 times wider in the $J=8$ case.

These results reflect a weak-coupling polaron structure that is more rigid than is expected by adiabatic theory in 1-D, but more correlated with phonons than is expected by adiabatic theory in 2-D and 3-D; spatially, weak-coupling polarons are much more compact than is expected by adiabatic theory in 2-D and 3-D, but may be larger or smaller than the "large polarons" expected by adiabatic theory in 1-D.

Rather than stumble over jargon associated with distinctions now evaporated, it would appear most sensible to regard the qualitatively-similar correlated electron-phonon ground states of the weak-coupling regime as large polarons regardless of lattice dimensionality.
Moreover, since we have demonstrated that "free" states do not exist as the minimum-energy states at weak coupling in {\it any} dimension, the notion of self-trapping as a change in the ground state from "free" to "self-trapped" character does not express the physical content of the phenomenon.
The self-trapping transition nonetheless exists as a clear feature of the polaron landscape.
In more operational and model-independent terms it is the more-or-less rapid transition from states characteristic of the weak-coupling regime to states characteristic of the strong-coupling regime as reflected in observable polaron properties; i.e., a transition from large polaron structure to small polaron structure.
Again contrary to adiabatic theory, which asserts the absence of a self-trapping transition 1-D because of the absence of free-electron ground states, this notion of a self-trapping transition {\it does} exist in 1-D as well as in 2-D and 3-D.

It is not the case, of course, that adiabatic theory is everywhere invalid.
It appears from multiple considerations, however, that part of what occurs in the course of the self-trapping transition is a dematerialzation of the locking relation between electron and phonon coordinates that is essentially universal in adiabatic approaches \cite{Brown97a}.
On the strong-coupling side of the transition, this locking is endemic and semi-classical treatments appear to enjoy respectable agreement with fully-quantum mechanical strong-coupling perturbation theory; on the weak coupling side, however, while still ultimately local in character as reflected in Figure~\ref{fig:corr123d}, electron-phonon correlations are increasingly sensitive to the discrete quantization of phonon energies and momentum-space structure.

These finer details are not well-captured quantitatively by the Merrifield method, but a very accurate and detailed description is available in the Global-Local variational method that follows from direct, sequential generalization of the Merrifield method.
We have used the Global-Local method to analyze 1-D polaron structure in detail throughout the polaron parameter space\cite{Romero98a,Brown97b}, and higher-dimensional calculations by this method are in progress\cite{Romero98}.

\section*{Acknowledgement}

The authors gratefully acknowledge P. E. Kornilovitch and G. Kalosakas for providing numerical data used in Figure~\ref{fig:ground}.
This work was supported in part by the U.S. Department of Energy under Grant No.
 DE-FG03-86ER13606.

\bibliography{../Bibliography/theory,../Bibliography/books,../Bibliography/experiment,../Bibliography/temporary}

\begin{thebibliography}{10}

\bibitem{Toyozawa61}
Y. Toyozawa, Prog. Theor. Phys. {\bf 26},  29  (1961).

\bibitem{Toyozawa63}
Y. Toyozawa,  in {\em Polarons and Excitons}, edited by C.~G. Kuper and G.~D.
  Whitfield (Plenum, New York, 1963), p.\ 211.

\bibitem{Emin73}
D. Emin, Adv. Phys. {\bf 22},  57  (1973).

\bibitem{Sumi73}
A. Sumi and Y. Toyozawa, J. Phys. Soc. Japan {\bf 35},  137  (1973).

\bibitem{Emin76}
D. Emin and T. Holstein, Phys. Rev. Lett. {\bf 36},  323  (1976).

\bibitem{Ueta86}
M. Ueta {\it et~al.}, {\em Excitonic Processes in Solids} (Springer-Verlag,
  Berlin, 1986).

\bibitem{Kabanov93}
V.~V. Kabanov and O.~Y. Mashtakov, Phys. Rev. B {\bf 47},  6060  (1993).

\bibitem{Toyozawa80a}
Y. Toyozawa and Y. Shinozuka, J. Phys. Soc. Jap. {\bf 48},  472  (1980).

\bibitem{Silinsh94}
E.~A. Silinsh and V. {\v{C}\'{a}pek}, {\em Organic Molecular Crystals:
  Interaction, Localization and Transport Phenomena} (AIP-Press, New York,
  1994).

\bibitem{Song96}
K.~S. Song and R.~T. Williams, {\em Self-Trapped Excitons} (Springer Verlag,
  Berlin, 1996).

\bibitem{Toyozawa97}
Y. Toyozawa, Pure \& Appl. Chem. {\bf 69},  1171  (1997).

\bibitem{Merrifield64}
R.~E. Merrifield, J. Chem. Phys. {\bf 40},  445  (1964).

\bibitem{Zhao97a}
Y. Zhao, D.~W. Brown, and K. Lindenberg, J. Chem. Phys. {\bf 106},  5622
  (1997).

\bibitem{Romero98a}
A.~H. Romero, D.~W. Brown, and K. Lindenberg, to appear in J. Chem. Phys.
  (1998).

\bibitem{Holstein59a}
T. Holstein, Ann. Phys. (N.Y.) {\bf 8},  325  (1959).

\bibitem{Holstein59b}
T. Holstein, Ann. Phys. (N.Y.) {\bf 8},  343  (1959).

\bibitem{Alexandrov95}
A.~S. Alexandrov and S.~N. Mott, {\em Polarons \& Bipolarons} (World
  Scientific, London, 1995).

\bibitem{Romero98e}
A.~H. Romero, D.~W. Brown, and K. Lindenberg, cond mat,  9809025  (1998).

\bibitem{Romero98c}
A.~H. Romero, D.~W. Brown, and K. Lindenberg, cond mat  9806031  (1998).

\bibitem{Romero98d}
A.~H. Romero, D.~W. Brown, and K. Lindenberg, cond mat  9808348  (1998).

\bibitem{Kalosakas98}
G. Kalosakas, S. Aubry, and G.~P. Tsironis, Phys. Rev. B {\bf 58},  3094
  (1998).

\bibitem{Kalosakas97}
G. Kalosakas, private communication  (1997).

\bibitem{Kornilovitch98a}
P.~E. Kornilovitch, cond-mat 9808155  (1998).

\bibitem{Kornilovitch98b}
P.~E. Kornilovitch, private communication  (1998).

\bibitem{Brown97a}
D.~W. Brown and K. Lindenberg, Physica D {\bf 113},  267  (1998).

\bibitem{Brown97b}
D.~W. Brown, K. Lindenberg, and Y. Zhao, J. Chem. Phys. {\bf 107},  3179
  (1997).

\bibitem{Romero98}
A.~H. Romero, D.~W. Brown, and K. Lindenberg, in preparation  (1998).

\end{thebibliography}

\end{document}